# Results from 3D Electroweak phase transition simulations

K. Farakos[a], K. Kajantie[b], M. Laine[b], K. Rummukainen[c*] and M. Shaposhnikov[d]

[a]National Technical University of Athens, Physics Department, Zografou, Gr-15780, Athens, Greece

[b]Department of Physics, P.O.Box 9, 00014 University of Helsinki, Finland

[c]Indiana University, Department of Physics, Swain Hall-West 117, Bloomington, IN 47405, USA

[d]Theory Division, CERN, CH-1211 Geneva 23, Switzerland

We study the phase transition in SU(2)-Higgs model on the lattice using the 3D dimensionally reduced formalism. The 3D formalism enables us to obtain highly accurate Monte Carlo results, which we extrapolate both to the *infinite volume* and to the *continuum* limit. Our formalism also provides for a well-determined and unique way to relate the results to the perturbation theory. We measure the critical temperature, latent heat and interface tension for Higgs masses up to 70 GeV.

## 1. WHY 3D SIMULATIONS?

Perturbative calculations have been extremely succesful in describing the physics of Electroweak interactions at zero temperature. However, at finite temperatures a purely perturbative analysis fails because of infrared problems: it is well known that the effective potential of the scalar field cannot be computed perturbatively for small $\phi$, in the symmetric phase. Thus, the calculation of the quantities characterizing the phase transition – for example, the critical temperature $T_c$, interface tension $\sigma$, and latent heat $L$ – requires the use of non-perturbative methods.

A direct way to include the non-perturbative effects is to perform 4D finite-temperature lattice simulations of SU(2)-Higgs models. However, in the interesting parameter range the theory is still weakly coupled, and we can use perturbative *dimensional reduction* (DR) to convert the 4D action into a 3D effective one. This step consists of integrating out all the massive modes (not constant in imaginary time) of the theory. In this talk we present results from 3D simulations with Higgs masses up to 70 GeV (for earlier results, see [1,2]; the results presented here will be described in detail in [3]).

We maintain that, in practice, *3D simulations are the method of choice for studying the EW phase transition* [4,5]:

**(I)** 3D model has one or two essential mass scales less than the original 4D model: in 4D, the lattice spacing $a$ and the linear size of the lattice $N_x$ have to satisfy the limits $T \ll a^{-1} \ll m_H(T)N_x$. In 3D, the heavy $T$-scale does not exist, and we have to require only that $m_W(T) \ll a^{-1} \ll m_H(T)N_x$.
**(II)** 3D theory is *superrenormalizable* — this gives an exact relation between the 3D lattice and continuum couplings in the limit $a \to 0$, and we can relate any lattice observable to the physical one for given Higgs and $W$ masses.
**(III)** For a given $a$ and $N_x$, the number of lattice variables is much less in 3D than in 4D, making the simulations easier.
**(IV)** We can consistently include the effects of *fermions* and even typical extensions of the Standard Model (for example, minimal SUSY extensions, the two-Higgs model) to the purely bosonic 3D SU(2)-Higgs simulations [5].

The dimensionally reduced 3D SU(2)-Higgs Lagrangian is formally similar to the 4D one:

$$L = \frac{1}{4}F^2 + (D_i\phi)^\dagger(D_i\phi) + m_3^2\phi^2 + \lambda_3(\phi^2)^2 \quad (1)$$

where $\phi^2 = \phi^\dagger\phi$ and the 3D couplings $g_3^2$ and $\lambda_3$ have dimension GeV (Here we discuss only the case where $A_0$ — the temporal component of the gauge field — is integrated over). We relate

*Presented by K. Rummukainen

the 3D couplings to the 4D ones at 2-loop level by Green's function matching [4,5]; using this method the nonlocal 2-loop terms which plaque straightforward DR [6] do not appear at all. The Lagrangian (1) is an approximation of the exact 3D one; by systematically estimating the effects of the neglected terms we can conclude that for $m_H \gtrsim 60\,\mathrm{GeV}$ the errors are less than 1%, depending on the observable.

The 3D lattice action can be written as

$$
\begin{aligned}
S &= \beta_G \sum_{x,i<j} (1 - \frac{1}{2}\mathrm{Tr}\, P_{x,ij}) \\
&- \beta_H \sum_{x,i} \frac{1}{2}\mathrm{Tr}\, \phi_x^\dagger U_{x,i} \phi_{x+i} \qquad (2) \\
&+ \sum_x [\phi^2 + \beta_R(\phi^2 - 1)^2].
\end{aligned}
$$

Due to superrenormalizability (II), we have an exact relation between lattice and continuum parameters $(\beta_G, \beta_H, \beta_R) \leftrightarrow (g_3^2 a, \lambda_3/g_3^2, m_3^2/g_3^4)$ when $a \to 0$; for example, $\beta_G = 4/(g_3^2 a)$ directly connects the coupling constant $\beta_G$ to the lattice spacing $a$. In 4D, the corresponding relation contains the RG constant $\Lambda_{\mathrm{Latt}}$, which has to be fixed by measurements. The 3D parameters are parametrized as $(h = m_H/80.6\,\mathrm{GeV})$ [3,5]

$$
\begin{aligned}
g_3^2 &= 0.44015\,T \\
\lambda_3/g_3^2 &= -0.00550 + 0.12622\,h^2 \qquad (3) \\
m_3^2/g_3^4 &= 0.39818 + 0.15545\,h^2 \\
&- 0.00190\,h^4 - 2.58088\,m_H^2/T^2\,.
\end{aligned}
$$

Note that neither $m_H$ nor $T$ here are true physical quantities. The values $m_H = 35$, 60 and 70 GeV used here correspond to physical pole masses $m_H(T{=}0) = 29.1$, 54.4 and 64.3 GeV in 4D SU(2)-Higgs theory (without fermions). For some other 4D theory (but the same 3D one, see (IV) above) the physical masses would be different [3]. Due to their transparency and universal nature we discuss only the 3D-values in the rest of the paper.

## 2. SIMULATIONS AND RESULTS

We fix the parameters according to eqs. (3) and use $m_H = 35$, 60 and 70 GeV. For each $m_H$, we use $\beta_G = 5$, 8, 12 and 20, which correspond to different lattice spacings. For each $\beta_G$ we have several volumes, allowing us to extrapolate the measurements (**A**) to the thermodynamical limit $V \to \infty$ and (**B**) to the continuum limit $a \to 0$. All in all, we have 59 different combinations of $m_H, \beta_G, V$.

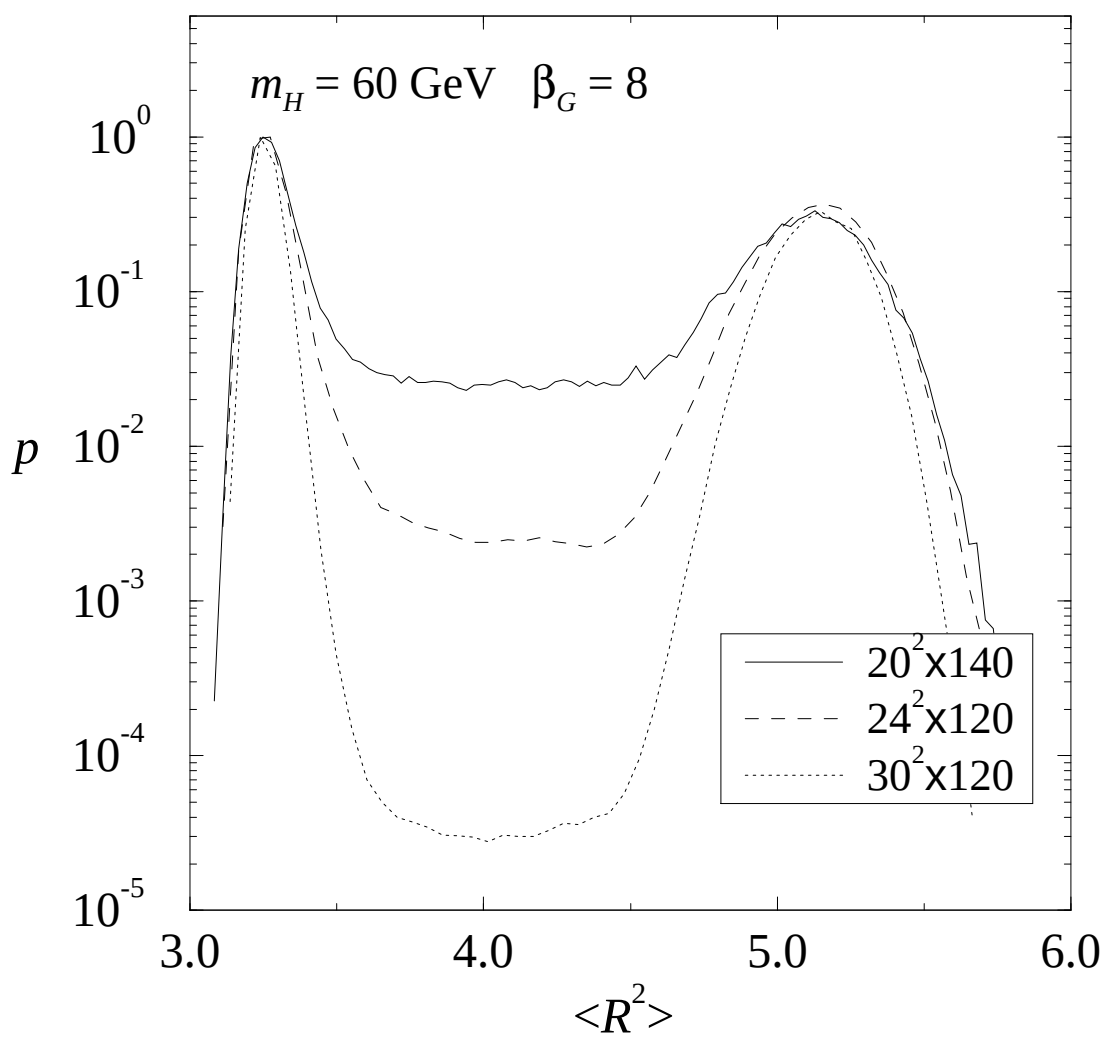

Figure 1. The distribution of $R^2 = \phi^\dagger\phi$ for some $m_H = 60\,\mathrm{GeV}$, $\beta_G = 8$ volumes.

For each lattice we search for the transition point by adjusting $\beta_H$. We have mainly concentrated our effort to the $m_H = 60\,\mathrm{GeV}$ case. In fig. 1 we show the distribution of $R^2 = \phi^\dagger\phi$ for the largest $\beta_G = 8$ volumes at the critical coupling $\beta_{H,c}$. The first order nature of the transition is obvious. As a rule, our 3D results qualitatively agree with the 4D results [7–9] and the recent 3D simulation [10]. However, the statistical errors in 3D are considerably smaller. For reviews, see [11,12].

**The critical temperature**

We monitor the phase transition with order parameters $R^2$ and $L = \frac{1}{3V}\sum_{x,i}\frac{1}{2}\mathrm{Tr}\,V_x^\dagger U_{x,i} V_x$, where $V$ is the SU(2) direction of the Higgs field $\phi = RV$. The critical coupling $\beta_{H,c}$ is located with several different methods:

(1) maximum of $C(L) = \langle (L - \langle L\rangle)^2\rangle$
(2) maximum of $C(R^2) = \langle (R^2 - \langle R^2\rangle)^2\rangle$
(3) minimum of the Binder cumulant of $L$:
$B(L) = 1 - \langle L^4\rangle/(3\langle L^2\rangle^2)$
(4) Equal weight value for the distribution $p(R^2)$
(5) Equal height value for the distribution $p(L)$

For each individual volume, the definitions (1)–

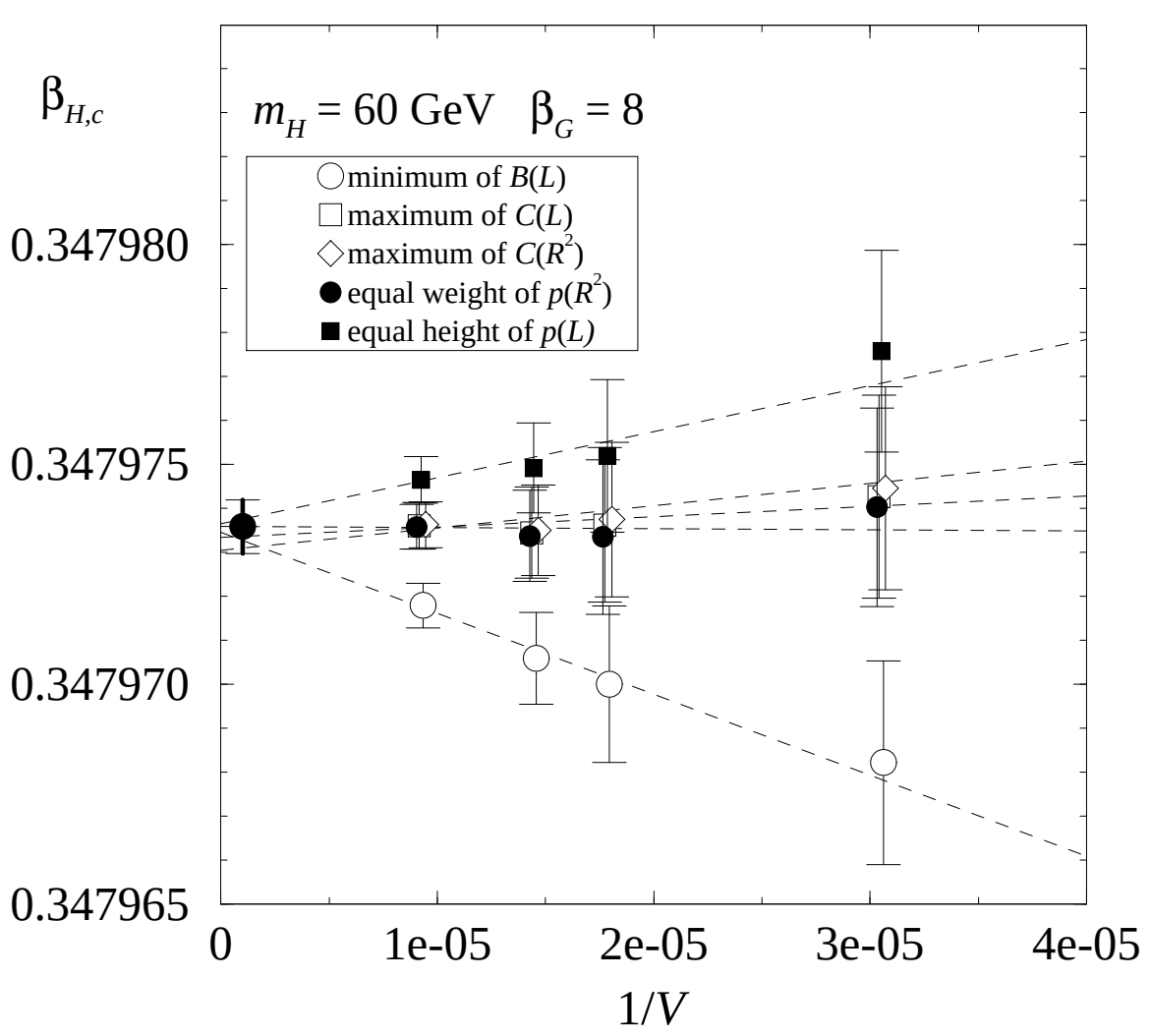

Figure 2. The $V \to \infty$ limit of $\beta_{H,c}$ measurements.

(5) yield different values for $\beta_{H,c}$, but when $V \to \infty$ all converge to the same limit, as is shown in fig. 2. For other values of $\beta_G$ the situation is similar.

For each $\beta_G$, we convert the $V = \infty$ value of $\beta_{H,c}$ to transition temperature $T_c$. These are in turn extrapolated to the continuum limit, as shown in fig. 3. For $m_H = 60$ we have high precision data for $\beta_G = 5$,8,12 and 20, and a good fit requires that we use a quadratic fit in $1/\beta_G$. For $m_H = 35$ and 70 GeV linear fits are acceptable. The final results are given in table 1; in all cases the transition is unambiguosly of first order.

Numerically, the $T_c$ values from the simulations are quite close to the perturbative ones, but due to the very high accuracy, they still differ at $\sim 10\sigma$ level, signaling significant non-perturbative and higher order perturbative effects.

Table 1
The critical temperature $T_c$, the interface tension $\sigma$ and the latent heat $L$ for different Higgs masses. The value of $\sigma$ at $m_H = 35$ GeV comes only from $\beta_G = 8$ simulations.

| $m_H$/GeV | 35 | 60 | 70 |
|---|---|---|---|
| $T_c$/GeV | 92.64(7) | 138.38(5) | 154.52(10) |
| $T_c^{\rm pert}$/GeV | 93.3 | 140.1 | 157.0 |
| $\sigma/T_c^3$ | [0.0917(25)] | 0.0023(5) | — |
| $\sigma^{\rm pert}/T_c^3$ | 0.061 | 0.008 | 0.005 |
| $L/T_c^4$ | 0.256(8) | 0.0406(7) | 0.0273(16) |
| $L^{\rm pert}/T_c^4$ | 0.22 | 0.041 | 0.028 |

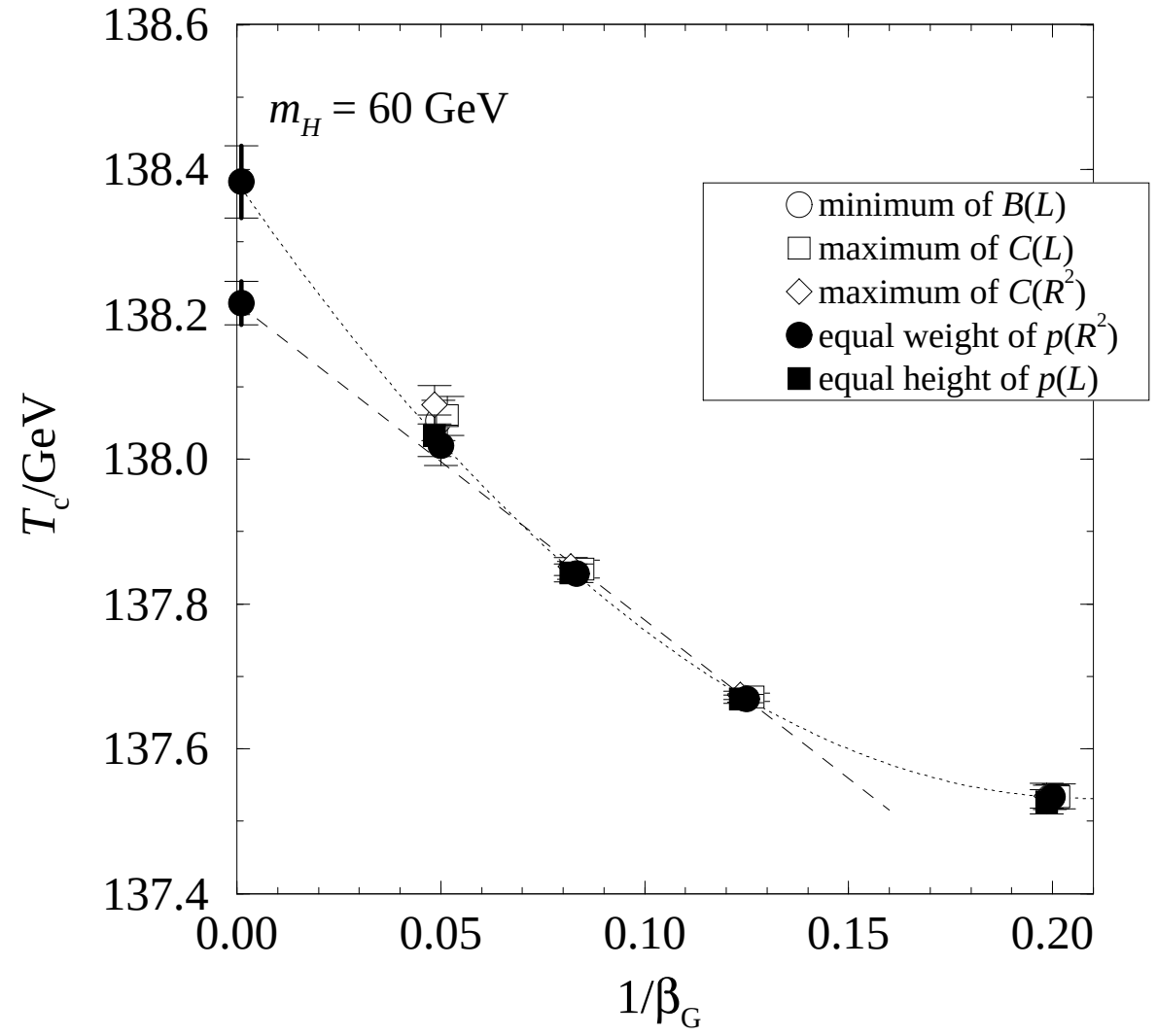

Figure 3. The continuum limit of the critical temperature $T_c$ for $m_H = 60$ GeV. Only the quadratic fit has an acceptable $\chi^2$/d.o.f.

**The interface tension and the latent heat**

We measure the interface tension with the *histogram method*: at the critical temperature the distribution of the order parameter develops a double-peak structure (fig. 1). The interface tension can be extracted from the limit

$$\frac{\sigma}{T} = \lim_{V \to \infty} \frac{1}{2A} \log \frac{P_{\rm max}}{P_{\rm min}} \,, \qquad (4)$$

where $A$ is the area of the interface and $P_{\rm max}$ and $P_{\rm min}$ are the distribution maximum and the minimum between the peaks. To use eq. (4) finite size corrections are needed; for details, see [3]. A crucial requirement is the "flat minimum" in the distribution between the peaks; this excludes all but the largest cylindrical volumes from the analysis.

In fig. 4 we show the $V \to \infty$ extrapolation of $\sigma$ for $m_H = 60$ GeV. These values are then further extrapolated to $\beta_G \to \infty$; the final value is $\sigma = 0.0023(5)\,T_c^3$. This is substantially smaller than the perturbative result $0.008\,T_c^3$, and signals the presence of non-perturbative effects for $\sigma$.

For $m_H = 35$ GeV we cite only $\beta_G = 8$ result (table 1), since we do not have "flat" histograms for other $\beta_G$ values; the continuum value is quite likely considerably smaller. For $m_H = 70\,\mathrm{GeV}$ we cannot reliably extract any non-trivial value.

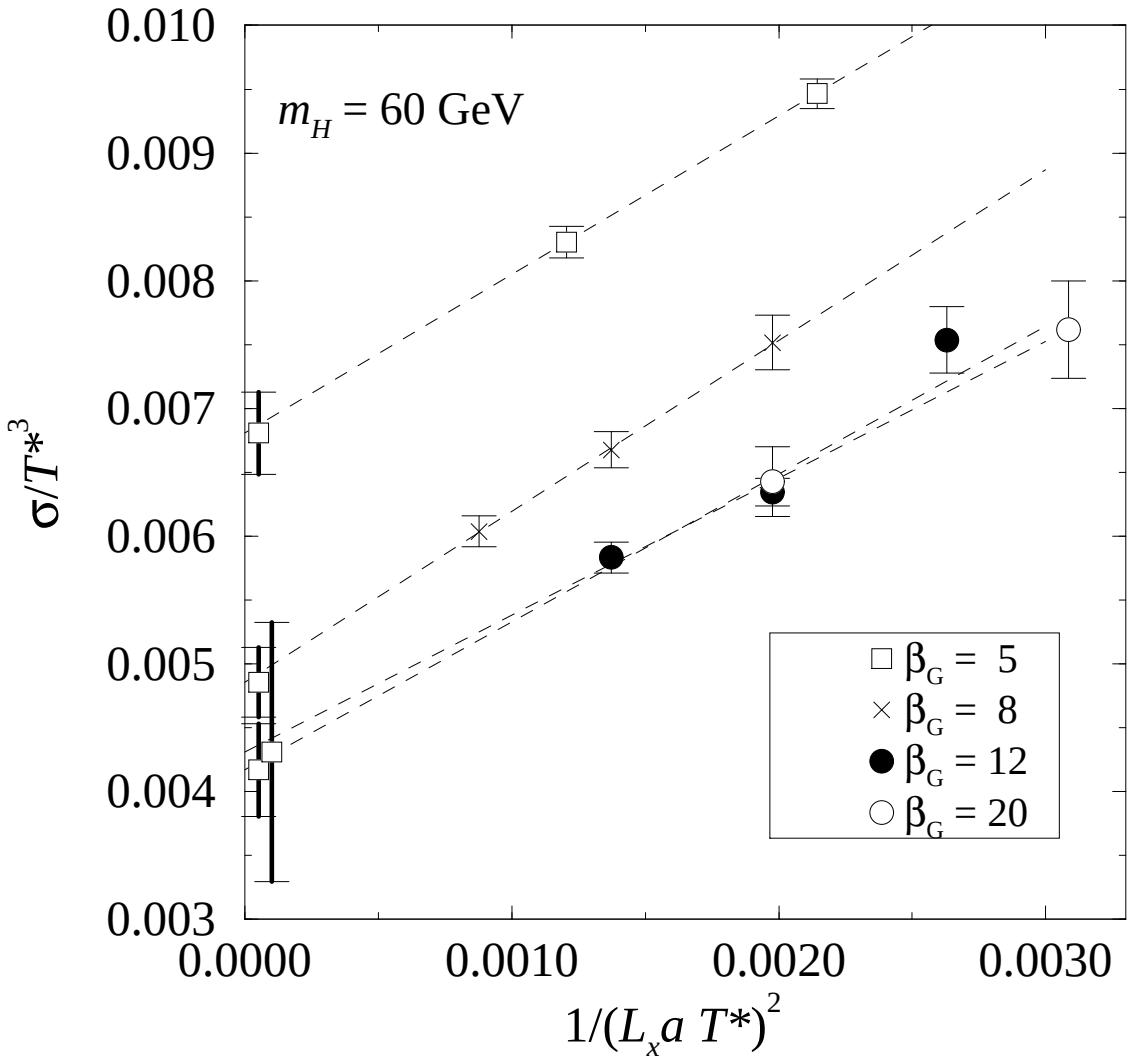

Figure 4. The interface tension extrapolated to $V \to \infty$ for $m_H = 60\,\mathrm{GeV}$.

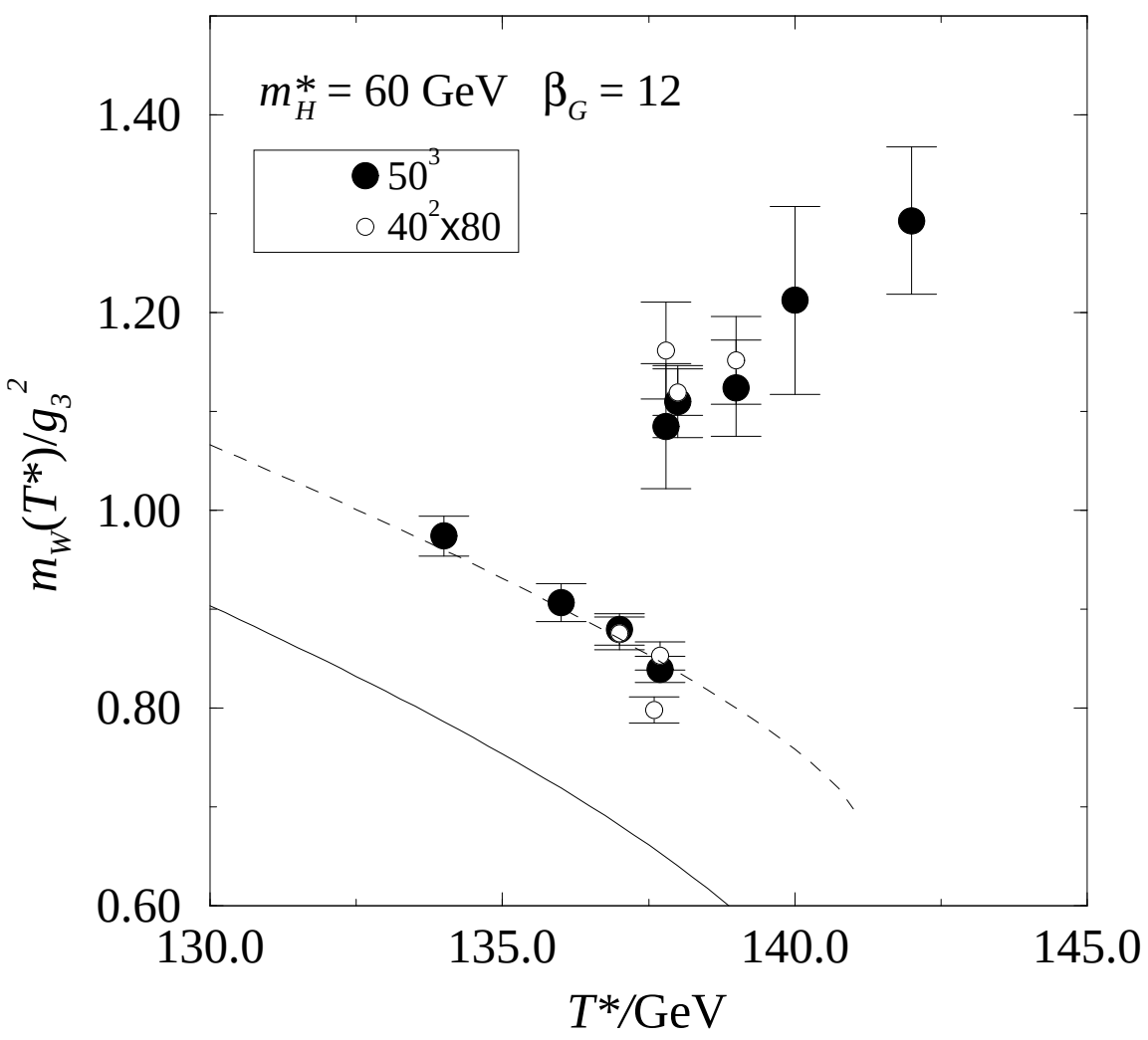

Figure 5. The $W$ mass around $T_c$ for $m_H = 60\,\mathrm{GeV}$ and $\beta_G = 12$.

The latent heat $L$ can be extracted from the discontinuity of $R^2$ at $T_c$. For details, we again refer to [3]; in contrast to $\sigma$, the continuum limit can be taken for all $m_H$, and the results are remarkably close to the perturbative values, as can be seen from table 1.

**The Higgs and $W$ masses**

In order to measure $m_H(T)$ and $m_W(T)$ we perform a separate series of simulations around $T_c$ for $m_H = 60\,\mathrm{GeV}$. We observe a good scaling between $\beta_G = 8$ and 12. Both $m_H(T)$ and $m_W(T)$ have a discontinuity at $T_c$, and the masses are *higher* in the symmetric phase. In fig. 5 we show $m_W(T)$ in units of $g_3^2 = g^2 T$. The value of $m_W(T > T_c)$ contradicts the analytical limit $m_W/g_3^2 \lesssim 0.29$ [13]. Similar behaviour has been observed in 4D [9] and 3D [10] simulations at smaller $m_H$.